\documentclass[conference]{IEEEtran}

\usepackage[utf8]{inputenc}
\usepackage{graphicx}
\usepackage{amsmath}
\usepackage{booktabs}
\usepackage{caption}
\usepackage{float}
\usepackage{cite}
\usepackage{hyperref}
\usepackage{enumitem}
\usepackage{listings}

\title{Evaluating Generative AI for CS1 Code Grading: Direct vs Reverse Methods}

\author{
    \IEEEauthorblockN{Ahmad Memon}
    \IEEEauthorblockA{
        University of British Columbia Okanagan \\
        Email: amemon03@student.ubc.ca
    }
    \and
    \IEEEauthorblockN{Abdallah Mohamed}
    \IEEEauthorblockA{
        University of British Columbia Okanagan \\
        Email: abdallah.mohamed@ubc.ca
    }
}

\begin{document}

\maketitle

\begin{abstract}

Manual grading of programming assignments in introductory computer science courses can be time-consuming and prone to inconsistencies. While unit testing is commonly used for automatic evaluation, it typically follows a binary pass/fail model and does not give partial marks. Recent advances in large language models (LLMs) offer the potential for automated, scalable, and more objective grading. 

This paper compares two AI-based grading techniques: \textit{Direct}, where the AI model applies a rubric directly to student code, and \textit{Reverse} (a newly proposed approach), where the AI first fixes errors, then deduces a grade based on the nature and number of fixes. Each method was evaluated on both the instructor's original grading scale and a tenfold expanded scale to assess the impact of range on AI grading accuracy. To assess their effectiveness, AI-assigned scores were evaluated against human tutor evaluations on a range of coding problems and error types.

Initial findings suggest that while the Direct approach is faster and straightforward, the Reverse technique often provides a more fine-grained assessment by focusing on correction effort. Both methods require careful prompt engineering, particularly for allocating partial credit and handling logic errors. To further test consistency, we also used synthetic student code generated using Gemini Flash 2.0, which allowed us to evaluate AI graders on a wider range of controlled error types and difficulty levels. We discuss the strengths and limitations of each approach, practical considerations for prompt design, and future directions for hybrid human-AI grading systems that aim to improve consistency, efficiency, and fairness in CS courses.

\end{abstract}


\section{Introduction}

Introductory computer science courses often enroll large numbers of students, creating significant grading workloads for instructors and teaching assistants (TAs). Manual grading is time consuming and, due to human variability, can exhibit inconsistencies, especially when awarding partial credit for partially correct logic or nuanced errors. These challenges highlight the need for scalable and fair grading systems.

One commonly adopted solution is the use of unit tests. However, unit testing frameworks come with two major limitations: first, they operate on a binary logic—either a solution passes or fails—and lack the nuance to assess partial correctness. Second, designing and validating comprehensive unit tests is itself a time-intensive process that must be repeated for every new assignment.

Recent advances in large language models (LLMs) and Generative AI (GenAI) open new possibilities for automated code assessment. These models can interpret student code, detect semantic and syntactic issues, and apply grading rubrics with increasing reliability. However, the feasibility of such systems hinges on whether AI can accurately mirror human judgments, particularly for complex or poorly structured submissions where partial credit is subtle. Our exploration builds on prior research in automated assessment but focuses on prompt engineering the AI for two distinct grading strategies.

\subsection{Scope and Contributions}

This paper proposes an AI-assisted grading pipeline (Fig.~\ref{fig:flowchart}) that complements existing unit testing frameworks in introductory CS courses. The system is used when code submissions fail unit tests and require more careful evaluation for partial credit. The aim is to improve grading consistency and scalability while preserving pedagogical fairness. Our contributions are as follows:
\begin{enumerate}
    \item We introduce and compare two AI-based grading strategies: the conventional Direct method, where a rubric is applied directly to student code, and a novel Reverse method, in which the AI first debugs the code and then assigns a score based on the type and number of corrections made.
    \item We conduct a controlled research study to evaluate and compare the effectiveness of Direct and Reverse grading methods. Using a range of student-like submissions categorized by quality (Poor, Moderate, Good), we compare AI-assigned scores to those given by experienced human TAs. This study is guided by four research questions:   
    \begin{enumerate}[label=\textbf{RQ\arabic*}:] 
        \item How accurate is AI grading compared to human tutors for CS1 code submissions, including both syntax and logic errors?
        \item How does the Reverse grading technique compare to Direct grading in terms of grading quality and feedback?
        \item How does rubric granularity (e.g., 10-point vs 100-point) affect grading accuracy across methods?
        \item Can AI systems reliably identify submissions where human review is needed due to low confidence or ambiguous logic?
    \end{enumerate}
\end{enumerate}

\begin{figure}[H]
\centering
\includegraphics[width=0.4\textwidth]{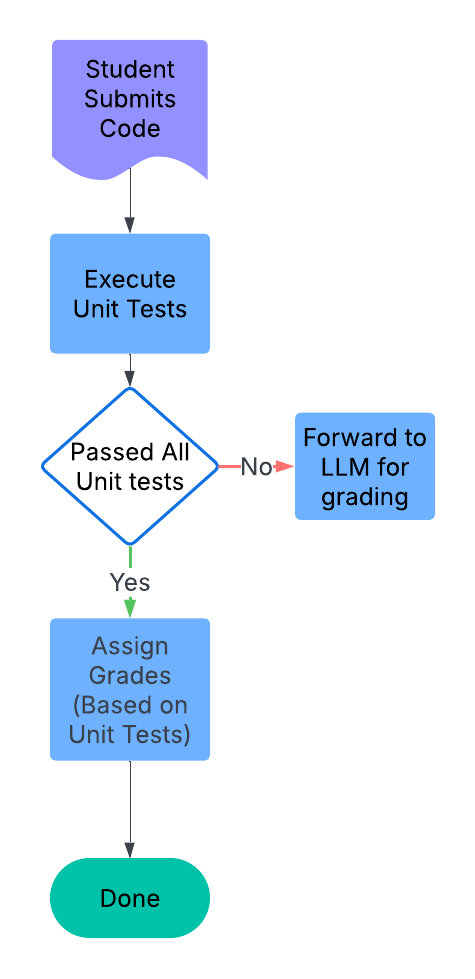}
\caption{
Proposed AI-assisted grading pipeline for CS1 coding assignments. Submissions are first evaluated using unit tests. If they pass, grades are assigned automatically. If they fail, the code is forwarded to a large language model (LLM) for deeper evaluation using either Direct (rubric-based) or Reverse (fix-then-grade) strategies. This design enables scalable grading while supporting nuanced assessment of partially correct solutions.
}
\label{fig.flowchart}
\end{figure}

\section{Related Work} \label{Related Work}

Large language models (LLMs) are being actively explored for their potential to support automated grading and feedback in educational settings, especially for programming and conceptual tasks.

Mohamed et al. \cite{Mohamed2025} conducted a comprehensive evaluation of LLMs including Gemini 1.5 Pro, GPT-4o, LLaMA 3 70B, and Mixtral for grading programming assignments. They found that some models align well with human graders but often struggle with complex logic or nuanced errors.

Pankiewicz and Baker \cite{PankiewiczBaker2023} used GPT-3.5 to generate personalized hints for debugging code. Their controlled experiment showed improved student performance and reduced time to fix errors, but also raised concerns about over-reliance on AI-generated hints.

Wan and Chen \cite{c} focused on conceptual feedback in physics education. By using few-shot and chain-of-thought prompting with GPT-3.5, they produced feedback rated as accurate and useful, though they recommended human validation for high-stakes grading.

Tobler’s Smart Grading system \cite{Tobler2023} allows instructors to provide sample solutions and rubrics that guide LLM evaluation of short-answer responses. The tool demonstrated consistent, structured feedback at scale.

Xie et al. introduced “Grade-Like-a-Human” \cite{Xie2024}, a multi-stage prompt-driven process involving rubric refinement, grading, and post-grading review. Their framework improved consistency and fairness by mimicking instructor grading practices.

Poličar et al. \cite{Policar2025} studied LLM-based grading in a bioinformatics course using both commercial (GPT-4, Nvidia 70B) and open-source models. They found that with well-designed prompts, open-source models delivered feedback comparable to human TAs.

While these studies highlight the promise of LLMs in educational grading, there remain important open questions. Few papers tackle grading partially correct or logically flawed code. There is also little exploration of how rubric resolution (e.g., 10- vs 100-point) affects evaluation, or how LLM confidence could help flag submissions for human review. These gaps directly motivate our work.

\section{Methodology}

\subsection{Dataset Creation}

We used synthetic programming questions designed to simulate typical CS1-level Java assignments. Each problem was paired with multiple student-like solutions representing different levels of correctness: Good, Moderate, and Poor. These submissions were generated using Gemini Flash 2.0, which allowed us to vary code quality in a controlled way and test the AI's ability to grade different types of mistakes. Collecting authentic student submissions required ethics approval and access to course data, which was not feasible within the study timeline. Therefore, we used synthetic submissions to ensure the evaluation remained systematic and reproducible. 
\begin{figure}[htbp]
\centering
\includegraphics[width=0.45\textwidth]{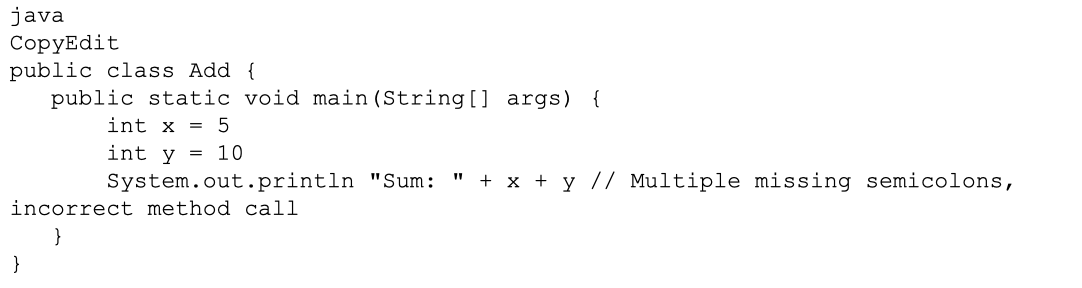}
\caption{Example of a synthetic \textit{Moderate} submission for the prime number task.}
\label{fig:moderate-solution}
\end{figure}

This submission contains a minor syntax issue (missing semicolon), representative of typical CS1-level errors. 
Such examples were graded by both AI and TAs, allowing us to compare alignment across quality levels.

\subsection{Grading Techniques}
\vspace{0.2em}

\subsubsection{Direct and Reverse Grading Methods}
\vspace{0.2em}

We evaluated two primary methods.

\begin{figure}[H]
\centering
\includegraphics[width=0.48\textwidth]{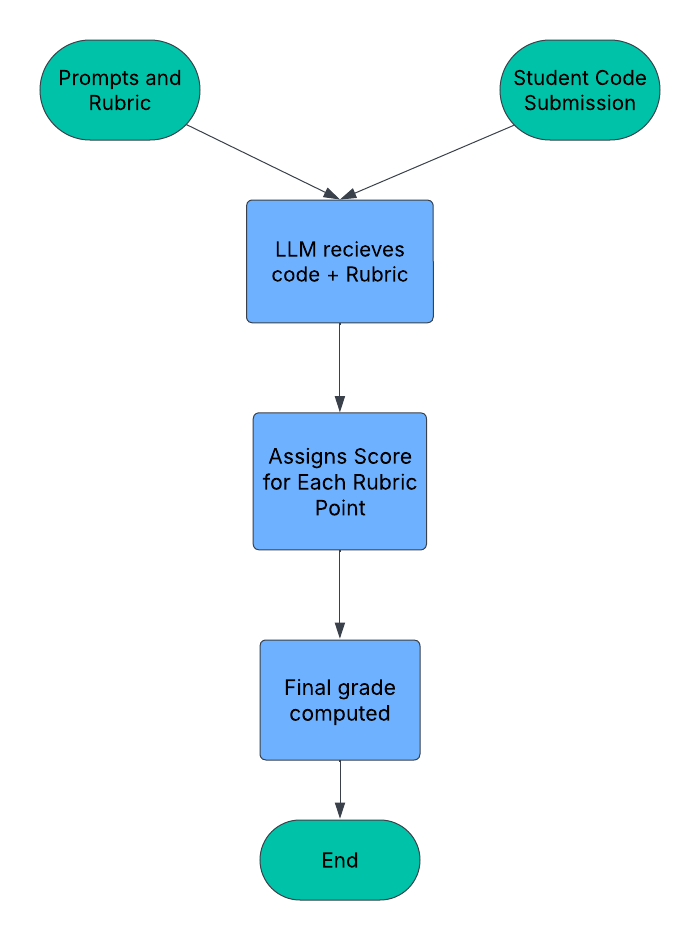}
\caption{Flowchart of the Direct Grading approach. The LLM receives the student's code along with a rubric, then scores each rubric category independently and outputs a rubric-aligned grade.}
\label{fig:direct}
\end{figure}

\textbf{Direct Grading}: As seen in Figure~\ref{fig:direct},  AI first receives the student code along with a human-designed rubric. It then applies the rubric point-by-point to assign a final grade. This approach mirrors the process a TA might follow when grading by hand.

\begin{figure}[H]
\centering
\includegraphics[width=0.6\textwidth]{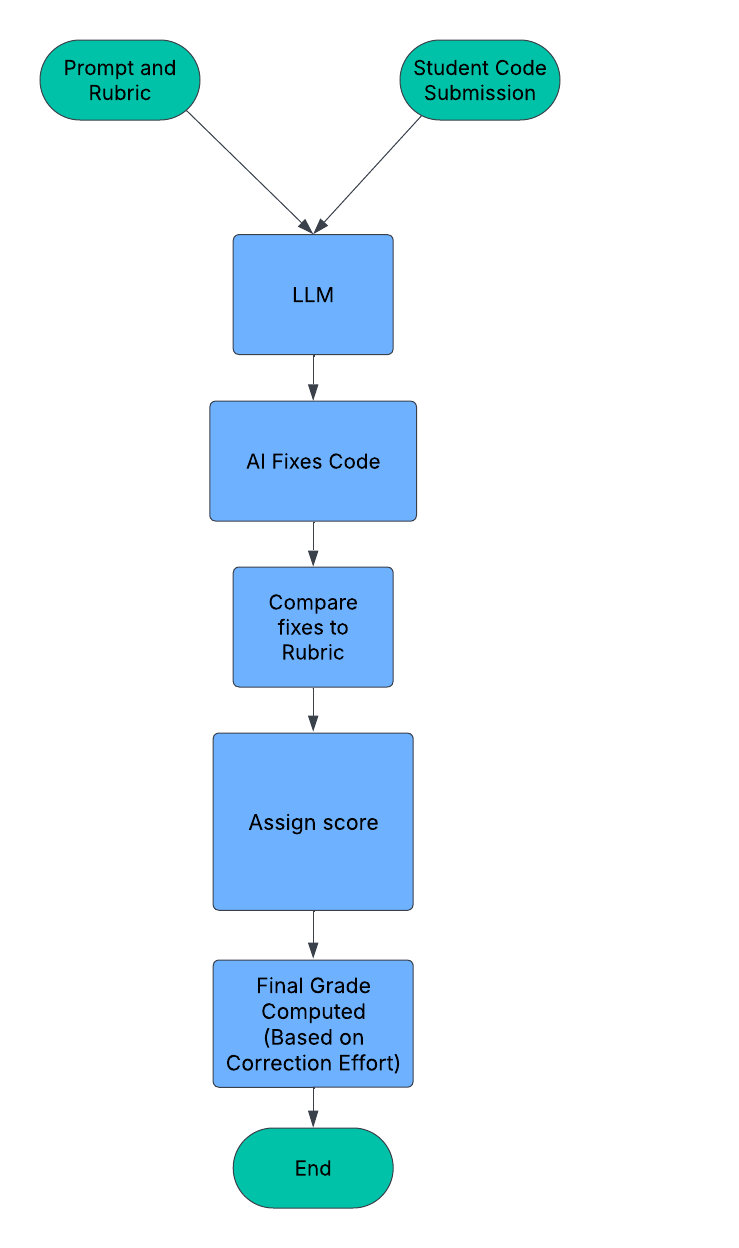}
\caption{Flowchart of the Reverse Grading approach. The LLM receives the student's code along with a rubric, but first attempts to fix the code. Based on the type and number of corrections made, it then uses the rubric to estimate a final score. This method emphasizes repair effort and debugging as part of the assessment.}
\label{fig:reverse}
\end{figure}

\textbf{Reverse Grading}: As seen in Figure~\ref{fig:reverse}, the AI first attempts to debug or correct the student submission. Then it estimates how many marks should be deducted based on the effort and complexity involved in correcting the errors. This method is intended to mimic how an instructor might assess "fixability" as part of awarding partial credit.

\vspace{1em}
\subsubsection{Rubric Scale}
We tested the grading prompts using two versions of the rubric: one with a 10-point scale and another with a 100-point scale. The 10-point rubric matched the format used by human TAs during course grading. To create the 100-point version, each rubric item was expanded to provide more fine grained scoring. This helped us explore whether AI could give more accurate or consistent grades when prompted with more detailed rubrics. These scaled prompts were only used with the AI and were not shown to human graders. When comparing results, scores from the 100-point system were scaled back down to match the original 10-point range.

\subsection{Prompt Engineering}

We iteratively refined the grading prompts for both Direct and Reverse techniques. Direct prompts instructed the AI to apply rubric criteria directly. Reverse prompts first asked the AI to correct the student code, summarize the fixes, and then assign a score. Full versions of these prompts are available in Appendix A.

\begin{quote}
“Score each rubric category out of 10. Explain each deduction in 1–2 lines. Then provide the final score and 1-line feedback.”
\end{quote}

In the Reverse method, the prompt structure asked the model to (1) fix the code, (2) explain the changes, (3) classify changes as minor or major, and (4) use this analysis to assign a final score with a reason. An example line from the final Reverse prompt:

\begin{quote}
“After fixing the code, estimate how many points should be deducted using the rubric. Subtract from a perfect score of 10.”
\end{quote}

Examples of final prompt versions for both methods are included in Appendix A.

\subsection{Evaluation and Human Baseline}

Human TAs independently graded the same set of code submissions using both 10-point and 100-point rubrics. Their scores served as the baseline for evaluating AI performance. We measured average absolute differences between human and AI scores.

\subsection{AI Model Selection }

We used two models in this study. GPT-4 was used to design and test the grading prompts for both the Direct and Reverse methods. We chose GPT-4 because earlier studies have shown that it works well for code-related tasks and gives clear, structured responses. Gemini Flash 2.0 was used to create synthetic student code examples at different quality levels. We did not compare the models directly, but used GPT-4 for grading because it was more reliable for prompt-based evaluation.
\section{Results and Findings}

\subsection{Score Comparison}

The table below shows the average scores assigned by each method, broken down by submission quality band. We analyze these scores in the following subsections, focusing on differences by quality band, grading method, and rubric scale.

\begin{table}[H]
\caption{Average Scores Across Methods (out of 10)}
\label{table:scores}
\centering
\begin{tabular}{@{}lccc@{}}
\toprule
Method & Poor & Moderate & Good \\
\midrule
Human TA               & 2.27 & 5.68 & 7.83 \\
Direct (10-pt)         & 3.20 & 6.12 & 7.54 \\
Direct (100-pt scaled) & 3.94 & 6.21 & 7.81 \\
Reverse (10-pt)        & 4.25 & 6.73 & 8.99 \\
Reverse (100-pt scaled)& 5.74 & 7.37 & 9.05 \\
\bottomrule
\end{tabular}
\end{table}

\subsection{Analysis of Coding Error Types}

We observed that syntax errors were consistently detected and scored reasonably across both approaches. Logic errors, especially repeated or compound logic mistakes, were more difficult for AI to penalize consistently.

Reverse grading generally performed better in identifying and correcting logical flaws before scoring. However, it occasionally missed repeated mistakes or provided similar feedback for logically different issues.

\subsection{Rubric Scaling}

We found that grading with a 100-point rubric allowed for more granular scoring. This helped both AI methods better approximate the TA scores. For example, reverse (100 scaled) often came within 0.5 points of TA grading on average.

In contrast, the 10-point rubric versions were more rigid. AI tended to cluster scores in round numbers (6, 7, 8) rather than capturing finer nuance. Thus, rubrics with higher resolution may be preferable when using AI grading.

\subsection{Prompt Refinement}
Prompt engineering was critical. In early versions, vague prompts often led to inconsistent or overly generous grade. Direct prompts benefited from explicit instructions to apply each rubric row independently and give concise rationales. Reverse prompts improved when we included the expectation to classify each fix (major/minor) and apply deductions accordingly.

As prompts evolved, grading consistency improved, especially for borderline Moderate / Good cases.

\begin{figure}[H]
\centering
\includegraphics[width=0.45\textwidth]{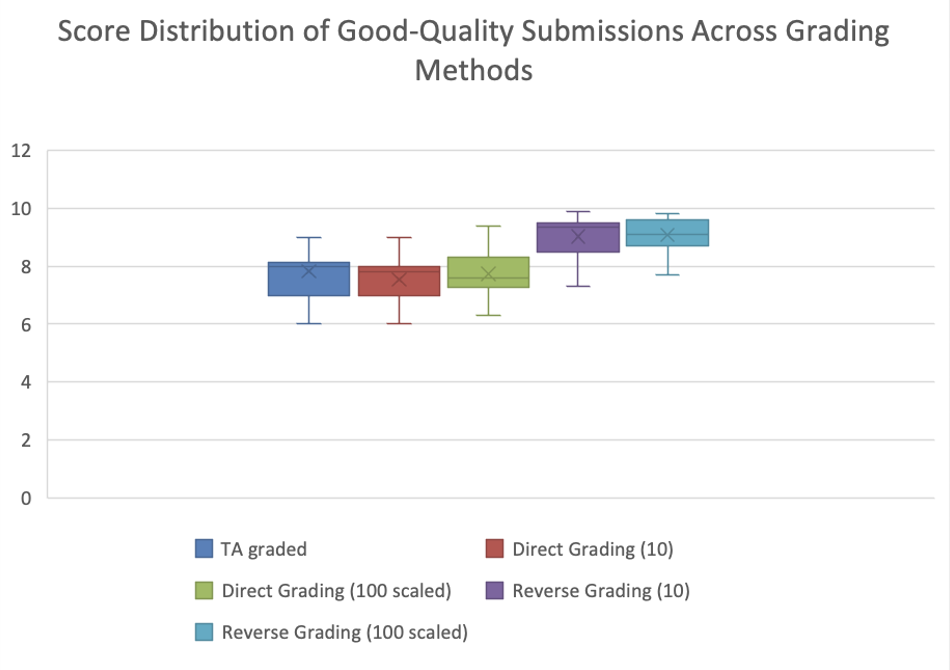}
\caption{Box plot: Good-quality submissions across grading methods. AI (especially Reverse methods) aligns closely with TA.}
\label{fig:good}
\end{figure}

\begin{figure}[H]
\centering
\includegraphics[width=0.45\textwidth]{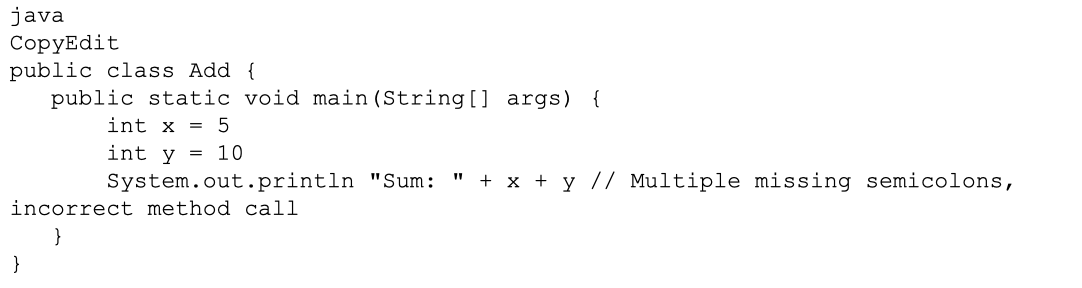}
\caption{Box plot: Moderate-quality submissions across grading methods. Reverse performs well, though spread is slightly higher.}
\label{fig:moderate}
\end{figure}

\begin{figure}[H]
\centering
\includegraphics[width=0.45\textwidth]{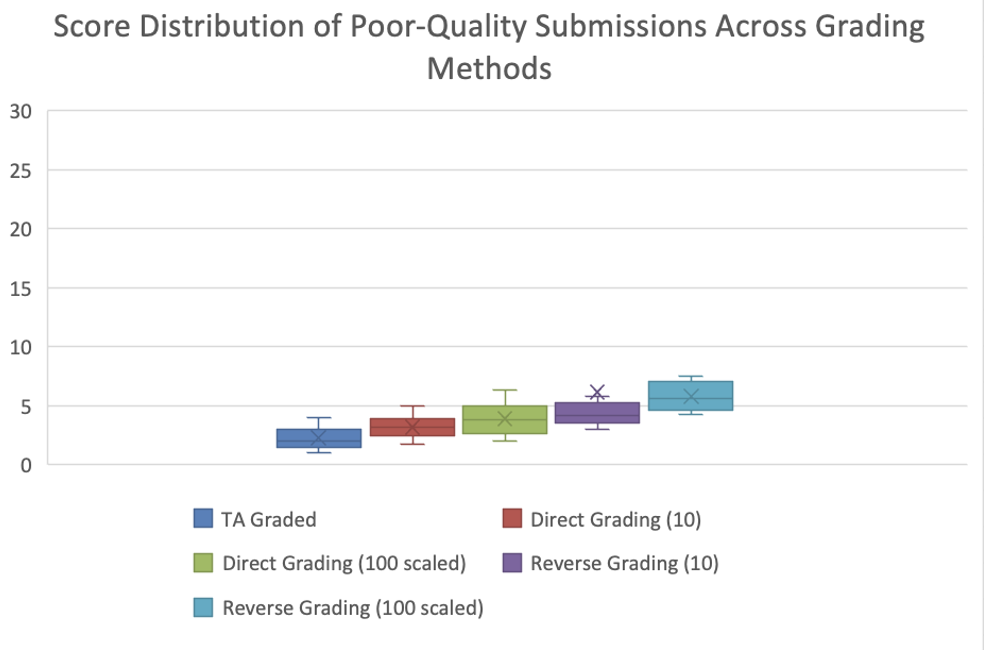}
\caption{Box plot: Poor-quality submissions across grading methods. Reverse scores tend to be higher than TA—may indicate under-penalization.}
\label{fig:poor}
\end{figure}

\begin{figure}[H]
\centering
\includegraphics[width=0.45\textwidth]{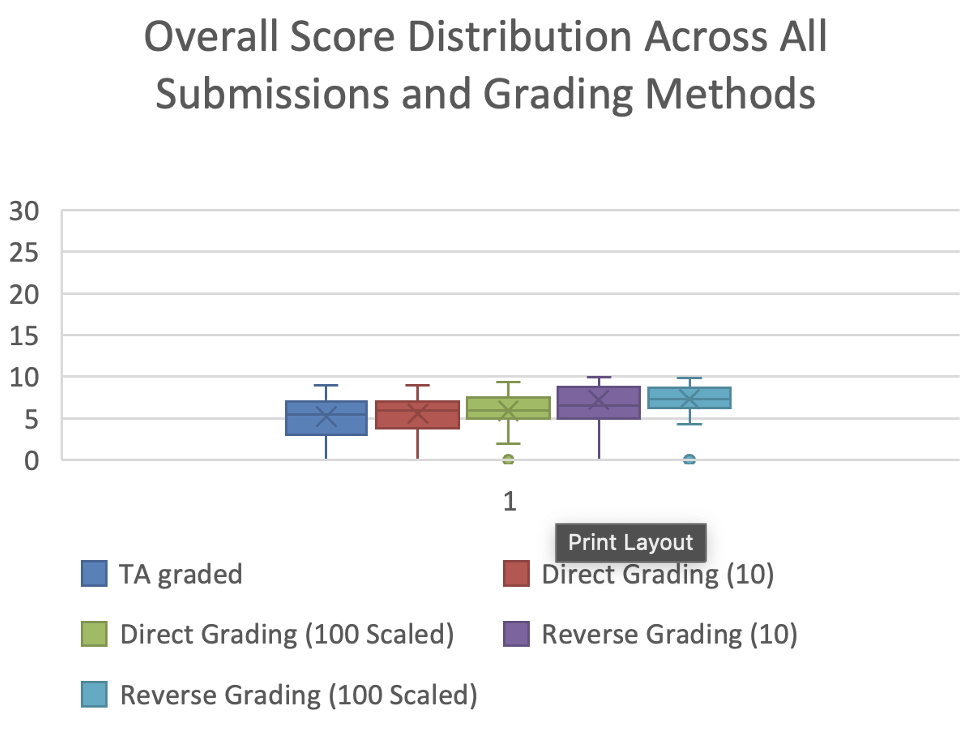}

\caption{Box plot: All submissions combined. AI shows promising consistency, though Reverse (100 scaled) skews slightly higher overall.}
\label{fig:all}
\end{figure}

To better visualize how the different grading methods performed across the three quality bands, we created a series of box plots, shown in Figures~\ref{fig:good} to~\ref{fig:all}.

\section{Discussion}

\subsection{Rubric Design}

One of the most important factors affecting model performance was rubric clarity. Rubrics that had specific, measurable items helped the model apply scores more consistently. In the Direct method, this helped the model stay focused on each category, while in the Reverse method, it guided how the model interpreted and explained the fixes it made. This highlights that rubrics do not only affect grading fairness for humans, but also directly shape the reasoning of automated systems. For instructors, it means that any AI grading setup needs to be paired with clear, detailed rubrics to avoid inconsistent or misleading feedback.

\subsection{Debugging vs. Direct Scoring}

The Direct approach is simpler to prompt and faster to run, but it sometimes failed to detect deeper logic issues, especially in submissions that looked structurally correct. The Reverse method offered a more detailed view of how the code could be repaired, and then scored based on the nature and number of corrections. This can have a positive impact on student learning, since the feedback shows not only what went wrong, but also how it could be fixed. In a real course, Reverse-style feedback might encourage students to reflect more on their mistakes and see grading as more constructive rather than purely judgmental. However, this method also requires more prompt engineering and computational time, so its practical use may depend on available resources.

While the overall box plots suggest that the Reverse method sometimes diverges from TA grading, this is largely due to its tendency to overestimate scores on poor-quality submissions. When analyzed by category, however, Reverse aligns more closely with TA grading on moderate and good submissions, which supports our earlier claim of its relative strength. This distinction between aggregate and per-category results explains the apparent contradiction.

\subsection{Hybrid Workflows and Human Review}

A combined workflow that uses AI for first-pass grading and instructors for review could be useful in large classes where time and consistency are concerns. This setup is especially valuable for submissions with unclear logic or cases where the model has low confidence. In those situations, forwarding the submission to a human reviewer helps maintain grading quality. In practice, such a system could ease the workload on teaching assistants while still giving instructors oversight where needed. It may also help maintain consistent grading across large sections or multiple graders. Although we did not implement a flagging system in this study, future work could explore how models might detect and mark uncertain cases for manual review. This kind of workflow supports scalable grading while keeping human judgment involved in edge cases, which may improve both fairness and student trust in the process.

\section{Limitations}

Although this work shows encouraging results, there are several limitations that must be considered. The code samples used in our study were short and focused on single-function problems. More complex programs, especially those involving multiple files or object-oriented design, may behave very differently. Our data set was also synthetic, created to simulate student submissions. Although this helped with consistency and allowed us to control the level of correctness, it does not capture the full range of styles, logic patterns, and mistakes that real students produce. This limits how well the results generalize to real classroom use. The models were guided using prompts, but were not specifically trained on our rubrics, and their output could vary with different instructions or model versions. The system may also struggle with edge cases, such as submissions that are technically correct but written in unusual or overly complex ways. Finally, there are technical concerns that were not explored in this study, including the cost of running large models, potential delays in response time, and fairness issues related to bias in AI-generated feedback. Taken together, these points show that while the approach has potential, it is still early work that needs further testing.

\section{Conclusion}

This study looked at how large language models can be used to grade programming assignments in CS1 courses. We compared two grading approaches: one that scores the code directly using a rubric, and another that first tries to fix the code and then scores it based on the changes made. Our results showed that both methods can come close to human grading, especially when using a detailed rubric and clear prompts. The fix-then-grade approach was often better at catching logic issues and giving more fine-grained feedback.

At the same time, this is early work with many limitations. We used synthetic data and short code examples, and the model was not trained directly on our grading standards. The system may not yet handle all types of submissions or more complex code. These results are promising, but they are just a first step. More testing is needed before this can be used in real classroom settings.

This study opens up several future directions:

Future work can include testing the grading methods on real student submissions collected over several semesters to see how they perform in everyday classroom settings. It would also be useful to try these methods on more advanced programming tasks, such as those that involve multiple functions or object-oriented design. Another direction is to add a system that marks submissions as uncertain when the model is not confident, so a human can take a second look. Future studies could also explore whether the type of feedback used in the Reverse method helps students understand their mistakes and improve. Finally, creating a shared collection of grading prompts and rubrics could make it easier for others to use and build on this work.

\appendices
\section{Appendix A: Prompt Samples}

\subsection{Direct Prompt}

Evaluate the following Java code using this rubric: Syntax, Logic, Output Correctness, and Style. Assign a score out of 10 for each category, and briefly explain any point deductions. At the end, provide the total score and a short summary of the student’s performance.

\subsection{Reverse Prompt}

Read the following Java code and try to correct it. List the fixes you made, and explain each one briefly. Then, estimate a score out of 10 based on how many and what kind of corrections were needed, using this rubric: Syntax, Logic, Output Correctness, and Style. Include the total score and a brief reason for the grade.

\end{document}